# Dual-color Coherent Perfect Absorber

Boyi Xue[1†], Jintian Lin[2†], Jiankun Hou[1], Yicheng Zhu[1], Ruixin Ma[1], Xianfeng Chen[3], Ya Cheng[2,4*], Li Ge[5], and Wenjie Wan[1,3*]

[1]State Key Laboratory of Advanced Optical Communication Systems and Networks,
University of Michigan-Shanghai Jiao Tong University Joint Institute, Shanghai Jiao Tong University, Shanghai 200240, China
[2]State Key Laboratory of High Field Laser Physics and CAS Center for Excellence in Ultra-Intense Laser Science, Shanghai Institute of Optics and Fine Mechanics (SIOM), Chinese Academy of Sciences (CAS), Shanghai 201800, China
[3]Department of Physics and Astronomy, Shanghai Jiao Tong University, Shanghai 200240, China
[4]School of Physics and Electronic Science, East China Normal University, Shanghai 200241, China
[5]Department of Physics and Astronomy, College of Staten Island, the City University of New York, NY 10314, USA

*† Equal contribution*
*Corresponding authors*: Wenjie Wan wenjie.wan@sjtu.edu.cn
or Ya Cheng  ycheng@phys.ecnu.edu.cn or Li Ge li.ge@csi.cuny.edu

**Perfect absorption of light critically affects light-matter interaction for various applications. Coherent perfect absorbers (CPA) gain the unique capability of controlling light with light in a linear fashion. Multi-color CPAs [Phys. Rev. Lett. 107, 033901] are highly desirable for broadband and nonlinear light-to-light coherent control, however, the experimental demonstration has still remained elusive. Here we experimentally observe a dual-color version of CPA (DC-CPA) through a second harmonic generation in a single whispering-gallery-mode microcavity. The DC-CPA enables simultaneous perfect absorption of both the incoming fundamental wave and its second harmonic. Similar to its linear counterpart, coherent control in the DC-CPA can be also realized by tuning the relative phase and intensity between the two-colored waves through nonlinear interference instead of the linear one. This scheme breaks the linear boundary of the traditional CPA into a multi-frequency domain and paves the way toward all-optically signal processing and quantum information.**

Coherent Perfect absorbers play a crucial role in various interdisciplinary areas encompassing optics [1,2], acoustics [4], microwaves [3], mechanics [5], as well as matter waves [6]. The ability of CPA to perfectly absorb any incoming wave without re-emitting solely lies on the interference-enhanced absorption in a lossy cavity, leading to a "blackhole" singularity in the spectrum [1]. Although initially proposed in optics, where coherent interferometers are straightforward to implement, similar ideas quickly span other coherent wave systems [1-6]. CPAs now have been realized on diverse platforms including cavity [7], disorder medium [8,9], metasurfaces [10,11], Parity-Time symmetric structures [12,13], and plasmonics [14]. Moreover, CPA promises a wide range of practical applications in optical information processing [15], photo-detection [16], sensing [17], and energy harvesting [18]. One critical quest of CPAs is to pursue the broadband/multi-channel perfect absorption across the spectra, aiming to achieve complete destructive interference over wide/multi-frequency channels. Prior works based on ultra-thin cavity design [14] or multi-layered structures [19] have been proposed for this purpose, alternatively, special wavefront shaping techniques can assist such multi-channel CPAs at a single frequency [9,20]. However, coherent control across multiple frequencies seems to be extremely hard to realize for CPAs [21,22].

Nonlinearity might be the key to tackling the problem, even though many prior works are still implemented in the linear regime. Contrarily, nonlinear waves' amplitude can greatly alter the dynamics of CPAs: a nonlinear version of CPA has been theoretically proposed [23], where the embedded nonlinear absorbing layer may prevent [6] or encourage the CPAs [24], even lead to a self-induced perfect absorption [25]. Interestingly, through nonlinear frequency conversion, effective scattering channels can be constructed between two or more frequency domains [26,27], such that, the interference can also occur between multiple frequencies, essential to the coherent control of CPAs [21]. Recently, through the second harmonic generation process, the nonlinear conversion introduces an additional nonlinear loss factor to break the critical coupling condition in a single-port CPA: the exact balance between the external coupling and the internal dissipation, this enables a self-induced transparency [25]. However, CPAs are still limited to the fundamental wave's frequency. Though "colored" CPAs in multi-frequency channels have been theoretically predicted through other nonlinear processes like optical parametric oscillation [21], the experiment still remains elusive till this work.

In this work, we demonstrate a dual-color version of coherent perfect absorption in a single whispering-gallery-mode microcavity. Contrasting to the conventional single-port CPA, this remarkable dual-color version enables simultaneous complete absorption of incoming waves at both the fundamental frequency and its second harmonics. The DC-CPA phenomenon arises from the nonlinear coupling between a fundamental wave (FW) optical mode at approximately 1560 nm and a second harmonic (SH) mode at around 780 nm. This results in coherent perfect destructive interference for both modes, similar to the critical coupling in the single-port CPA. All the incident energy is efficiently confined within the cavity, dissipating through material absorption and surface scattering mechanisms. Moreover, the coherent control of transmission can be realized by tuning the relative phase and intensity between these two input waves at two different frequencies. These results allow light-to-light controlled absorption, paving the way for all-optical information processing.

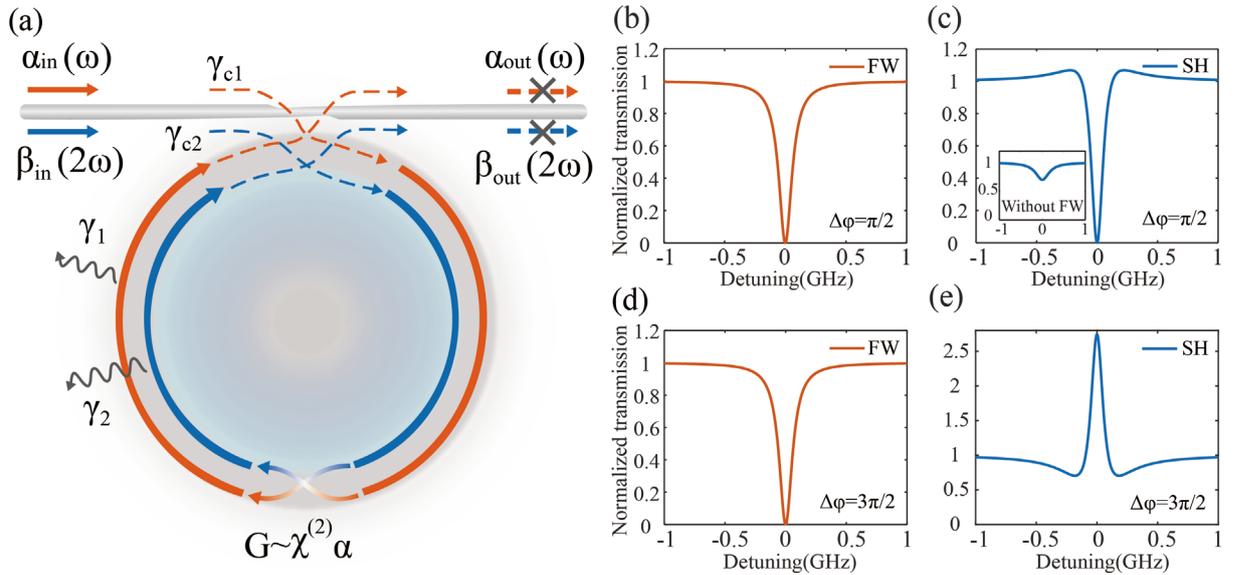

**Figure 1. Conceptual illustration of dual-color CPA effect.** (a) Schematic diagram of nonlinear coupling between two modes in a tapered fiber coupled microcavity. A strong pump field $\omega$ and a weak seed $2\omega$ are launch to the microcavity and coupled together through second-order nonlinearity. $\gamma_1$, $\gamma_2$: intrinsic loss for FW and SH respectively. $\gamma_{c1}$, $\gamma_{c2}$: coupling rate. G: the coupling coefficient which is proportional to $\chi^{(2)}$ and the intracavity FW field $\alpha$. (b) and (c) Their simulated transmission spectra under CPA condition with a phase difference of $\pi/2$. Inset in (c): Transmission spectrum for SH without FW. (d) and (e) Their simulated transmission spectra with a phase difference of $3\pi/2$.

A single-port CPA can be realized in a fiber-coupled microcavity scheme at the critical coupling condition [28], where zero output is achieved through destructive interference between the cavity's out-coupling light and residual transmission. Unfortunately, for the dual-color CPA, the critical coupling condition is contingent upon the input frequency. For instance, when adjusting the gap between the fiber and the cavity to achieve critical coupling of FW mode, the SH mode would possibly be under-coupled due to its shorter evanescent wave tail. Consequently, total absorption can only be attained in a single band in linear cases. However, dual-band critical coupling is possible in our DC-CPA scheme thanks to the nonlinear coupling between the two modes. We will illustrate later that the FW will remain critically coupled in the weak coupling regime, while the relatively weak SH wave will also become perfectly absorbed due to the total destructive interference among the out-coupled SH light, the nonlinearly converted SH, and the residual SH transmission [25].

The schematic illustration of the DC-CPA concept is depicted in Fig. 1. A second-order nonlinear microcavity, fabricated on a thin film Lithium Niobate [29], is coupled to a tapered fiber. Two input lasers are launched into the system, where $\alpha_{in}$ represents the amplitude of the input fundamental wave with frequency $\omega$, while $\beta_{in}$ denotes the input second harmonic wave at frequency $2\omega$. $\gamma_1$, $\gamma_2$ corresponding to the intrinsic decay rate for FW and SH respectively caused by material absorption and surface roughness, whereas $\gamma_{c1}$, $\gamma_{c2}$ represent the coupling strength to the single-mode tapered fiber. Due to negligible backscattering exhibited by this microcavity, only clockwise modes are excited at both wavelengths. Differing from the purely linear scenario, these two modes can also be coupled together due to second-order nonlinearity $\chi^{(2)}$. This process can be effectively described using the temporal coupled mode theory (TCMT). The nonlinear Hamiltonian of this system is expressed as

$$H_{eff} = \begin{pmatrix} \omega_1 - i\frac{\gamma_{c1}+\gamma_1}{2} & -2G^* \\ -G & \omega_2 - i\frac{\gamma_{c2}+\gamma_2}{2} \end{pmatrix}, \quad (1)$$

Where $\omega_1$, $\omega_2$ are the resonant frequencies for the intracavity fields of the FW and SH. We define the effective nonlinear coupling coefficient $G = g\alpha$, where $g$ represents the nonlinear coupling strength for second harmonic generation which is determined by $\chi^{(2)}$.

We adopt the S-matrix approach employed in the original one-port CPA [1], modify Eq. (1) [supplement S1], and re-describe our DC-CPA system utilizing the electromagnetic scattering matrix (S-matrix):

$$\begin{pmatrix} \alpha_{out} \\ \beta_{out} \end{pmatrix} = S(\omega) \begin{pmatrix} \alpha_{in} \\ \beta_{in} \end{pmatrix} = \begin{pmatrix} 1 - \frac{i\gamma_{c1}\delta_2}{\delta_1\delta_2 - 2|G|^2} & \frac{i\sqrt{\gamma_{c1}\gamma_{c2}}G}{\delta_1\delta_2 - 2|G|^2} \\ \frac{2i\sqrt{\gamma_{c1}\gamma_{c2}}G^*}{\delta_1\delta_2 - 2|G|^2} & 1 - \frac{i\gamma_{c2}\delta_1}{\delta_1\delta_2 - 2|G|^2} \end{pmatrix} \begin{pmatrix} \alpha_{in} \\ \beta_{in} \end{pmatrix}. \quad (2)$$

Where $\alpha_{out}$, $\beta_{out}$ are the output signals. We define $\delta_1 = \Delta_1 + i(\gamma_1 + \gamma_{c1})/2$ and $\delta_2 = \Delta_2 + i(\gamma_2 + \gamma_{c2})/2$, where $\Delta_1 = \omega - \omega_1$, and $\Delta_2 = 2\omega - \omega_2$ are the frequency detuning for FW and SH, respectively. For the sake of simplicity, we investigate the system under the condition of $\omega_2 = 2\omega_1$, such that their frequency detuning adheres to $\Delta_2 = 2\Delta_1$.

The eigenvalues and corresponding eigenvectors of the S-matrix can be found for any given $\omega$. If the eigenvalue approaches *infinity*, a non-zero output signal emerges even with infinitesimal input amplitudes, indicating the presence of a lasing mode. Conversely, a CPA mode occurs when the eigenvalue is *zero*, leading to the complete vanishing of the total output in response to an input signal aligned with the corresponding eigenvector of S. Notably, the nonlinear coupling strength $g$ in our experiment is significantly weak and the input power for the fundamental wave is also maintained at a low level. This ensures that the coupling coefficient G can be considered as a small constant value compared to other parameters ($\gamma_{1,2}$ and $\gamma_{c1,c2}$). Under this crucial condition, only the weak SH wave will be influenced during the nonlinear coupling process, while any alterations to the FW transmission spectrum are negligible. Consequently, this effectively precludes other nonlinear phenomena in the strong coupling regime such as self-induced transparency [25], and greatly simplifies the apparatus towards achieving DC-CPA through dual-port critical coupling.

The transmission spectra are numerically calculated and illustrated in Fig. 1(b)-(e). The theoretical expression of the normalized transmission is given by:

$$T_\omega = \left| 1 - \frac{i\gamma_{c1}\delta_2}{\delta_1\delta_2 - 2|G|^2} + \frac{i\sqrt{\gamma_{c1}\gamma_{c2}}G}{\delta_1\delta_2 - 2|G|^2}\frac{\beta_{in}}{\alpha_{in}} \right|^2, \quad (3)$$

$$T_{2\omega} = \left| 1 - \frac{i\gamma_{c2}\delta_1}{\delta_1\delta_2 - 2|G|^2} + \frac{2i\sqrt{\gamma_{c1}\gamma_{c2}}G^*}{\delta_1\delta_2 - 2|G|^2}\frac{\alpha_{in}}{\beta_{in}} \right|^2. \quad (4)$$

All the decay rate parameters are determined according to the experimental data. The FW is critically coupled, while the weak SH light is initially under-coupled in the absence of FW input, as depicted in the inset of Fig. 1c. Specific values for $G$ and $|\beta_{in}/\alpha_{in}|$ are selected in order to achieve the CPA condition, which will be further elucidated later. The transmission spectra of FW and SH are depicted in Fig. 1b, c, respectively, when their phase difference is $\pi/2$. The FW remains invariant since Eq. 3 will be reduced to the linear case under the weak coupling condition $G \ll \gamma_1, \gamma_2, \gamma_{c1}, \gamma_{c2}$. Meanwhile, the initially under-coupled SH is transformed into the critically coupling state. This can be readily comprehended in that the nonlinearity acts as an additional gain to balance $\gamma_2$ and $\gamma_{c2}$ ($\gamma_2 < \gamma_{c2}$ for under coupling), which results in the total destructive interference mechanism among three components: the intracavity SH light and the converted SH generated by FW are coupled out to destructively interfere with the residual SH transmission. Consequently, both waves are completely absorbed at zero detuning points, leading to the occurrence of the DC-CPA. In Fig. 1d, e, the relative phase is $3\pi/2$. The FW still remains unchanged while the SH spectrum shows an electromagnetically induced transparency (EIT) line shape whose transmission rate is above unity [30], which originates from the constructive interference among the three components mentioned above. These simulation results validate the existence of the DC-CPA effect and indicate its phase dependence.

According to the aforementioned theory, DC-CPA occurs at the zeros of the S-matrix. From Eq. 2, it is evident that the eigenvalues of the S-matrix are dependent on frequency $\omega$ and G, with $\omega$ being allowed to take any complex value. Since the corresponding $\omega$ at CPA zeros are located on the complex plane, it becomes crucial to manipulate the value of G by adjusting the input amplitude $\alpha_{in}$ in order to ensure a purely real number for $\omega$. The imaginary component of $\omega$ is plotted as a function of the square modulus of nonlinear coupling coefficient G in Fig. 2a. This result is numerically simulated with the same parameters employed in Fig. 1. Notably, $\omega$ is shifted onto the real axis when $G \approx 6.37 MHz$, which fulfills the DC-CPA condition.

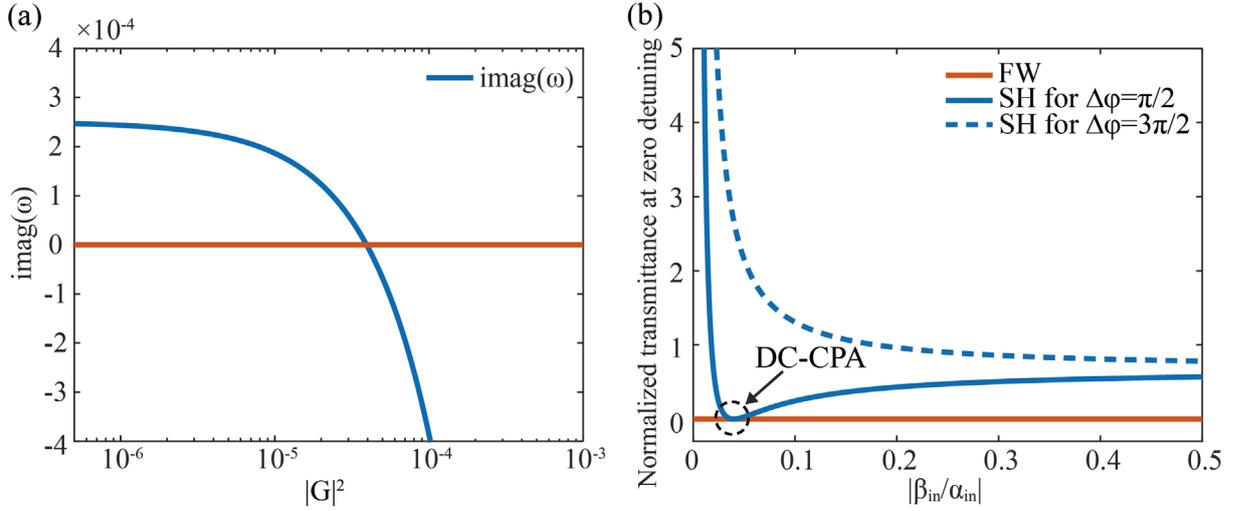

**Figure 2. Theoretical analysis of the DC-CPA conditions in a microcavity.** (a) The imaginary part of frequency $\omega$ for the zeros of the system's S-matrix is plotted as a function of $|G|^2$. The red line denotes $imag(\omega)=0$. The intersection point represents the CPA condition we require, where $G \approx 6.37MHz$. (b) Normalized transmittance of FW and SH modes at zero detuning points as a function of their relative input amplitude, which are calculated with different relative phases. The phase difference is $\pi/2$ and $3\pi/2$ for the solid line and dashed line, respectively. The results are obtained from the theoretical model described Eq. 3 and 4. The CPA condition determined in (a) is also utilized. The DC-CPA effect occurs when the input laser is launched into the eigenchannel of the S-matrix, as is marked in (b), where $\Delta\varphi = \pi/2$ and $|\beta_{in}/\alpha_{in}| = 0.0397$.

In addition to the requirement of processing zero eigenvalues, it is also imperative for the input modes to be launched into the eigen-channel defined by the eigenstate of S-matrix to achieve the DC-CPA phenomenon. This implies that the input fields $\alpha_{in}$ and $\beta_{in}$ must exhibit proper phase relationships as well as their amplitudes. The normalized transmittance of FW and SH at zero detuning point is depicted in Fig. 2b. It is evident that the perfect absorption only occurs for a phase difference of $\pi/2$. The transmittance of FW remains nearly constant, while SH exhibits significant variability under different relative amplitudes $|\beta_{in}/\alpha_{in}|$. When $|\beta_{in}/\alpha_{in}|$ approaches zero, the normalized transmittance of SH diverges, corresponding to the conventional second harmonic generation (SHG) process. As $|\beta_{in}/\alpha_{in}|$ increases, it initially decreases before eventually converging to the transmittance in the linear case. The DC-CPA effect could be observed at the point of zero transmittance, where $|\beta_{in}/\alpha_{in}| = 0.0397$.

In our experiment, a microcavity with a diameter of 200 microns and a high-quality factor of approximately $10^6$ is fabricated on a Lithium Niobate thin film. The microcavity is placed on a 3-axis nano-translation stage for precise positioning and coupled to a tapered fiber which is made of a standard single-mode fiber in the telecom band. A continuous laser source operating around 1560 nm wavelength is utilized to launch the FW mode, while its corresponding SH signal is produced by a second harmonic generator which contains a PPLN crystal. These signals are combined through a 780/1550 wavelength division multiplexer (WDM) and propagate inside the tapered fiber. The out-coupled signals are collected by the same fiber and separated using another WDM device. The FW signal is detected using a photodetector, whereas the SH mode is probed via a photomultiplier tube (PMT) with a response in the 780 nm band.

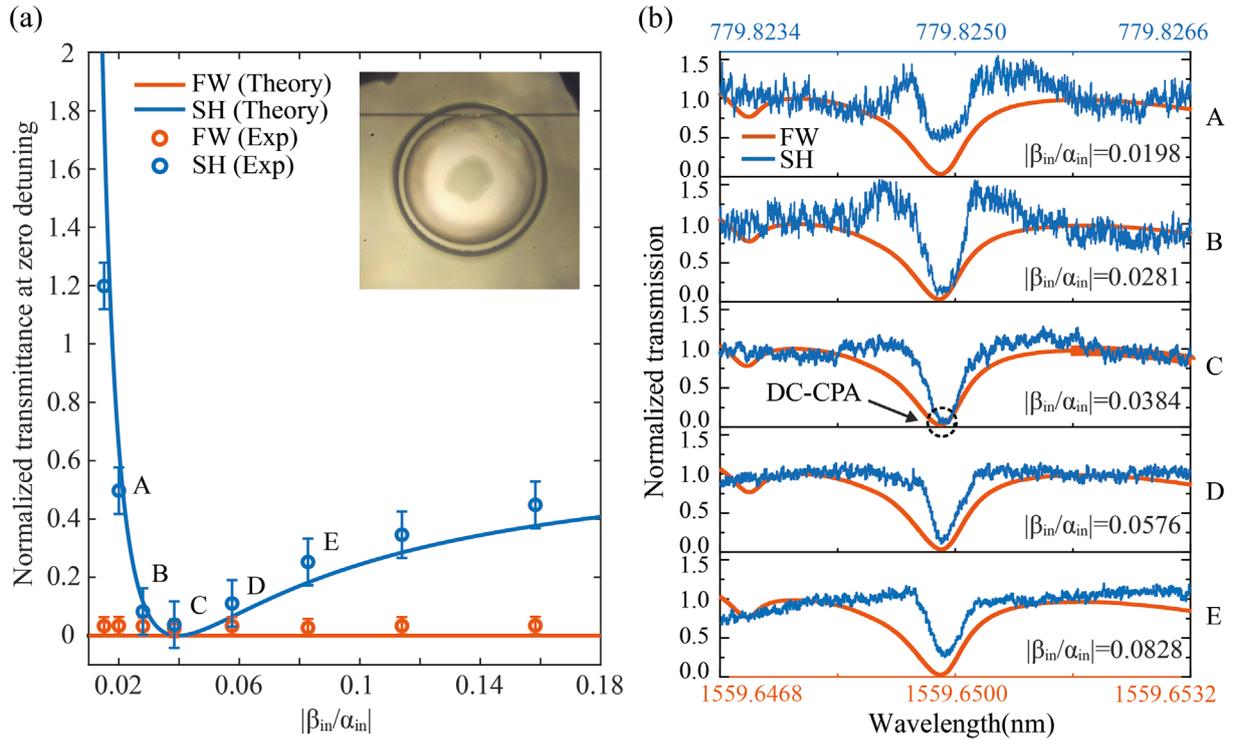

**Figure 3. Experimental observation of dual-color CPA.** (a) Normalized transmittance of FW and SH modes at zero detuning, which is plotted as a function of relative input amplitude $|\beta_{in}/\alpha_{in}|$. "Theory" refers to theoretical results, which are calculated with $G \approx 6.37 MHz$ and $\Delta\varphi = \pi/2$. While "Exp" denotes the experimental data, which is measured under approximately the same condition. Error bars represent the uncertainty in the measurement due to the system jitter. Inset in (a): Top-view picture of the microcavity device. (b) Experimentally obtained transmission spectra with different values of $|\beta_{in}/\alpha_{in}|$ corresponding to A~E in (a). notably, the DC-CPA point is marked in curve C, where the transmission rate is as low as 0.02599 and 0.03773 for FW and SH respectively.

We first launch the FW laser in isolation and select a mode suitable for generating a second harmonic wave inside the cavity. The relative position between the cavity and tapered fiber is precisely controlled using the nano-translation stage, ensuring approximate attainment of the critical coupling condition for the FW mode. In our experiment, the corresponding SH mode is observed to be under coupled conditions at this specific position, which can be attributed to the weaker evanescent tail associated with shorter wavelengths. The resonant frequencies are finely adjusted by tuning the cavity temperature using a thermoelectric cooler (TEC) to meet $\omega_2 = 2\omega_1$, ensuring that the condition of $\Delta_2 = 2\Delta_1$. The crucial condition of $G \approx 6.37 MHz$ can be obtained by meticulously tuning the amplitude of FW. However, in the experiment, it is challenging to determine its exact value. Consequently, we accomplish the DC-CPA effect by simultaneously adjusting the power of these two input modes through the attenuators while maintaining their relative phase fixed at $\pi/2$. Since the input power is varied at a low level (several dozens of microwatts for FW and several hundreds of nanowatts for SH), the transmission of FW remains constant due to weak nonlinear coupling as discussed previously. The power is meticulously adjusted until the outgoing signal of SH mode at zero detuning nearly vanishes. The DC-CPA phenomenon is observed when the input power levels of FW and SH are equal to $78.96 \mu W$ and $116.4 nW$ respectively, corresponding to a relative amplitude $|\beta_{in}/\alpha_{in}| = 0.0384$, which closely matches numerical simulations. The transmission spectrum is visualized in Graph C of Fig. 3b, revealing that the transmission rate approaches zero at the DC-CPA point, which can be further optimized by adjusting the coupling position and input power. Graphs A, B, D, and E in Fig. 3b represent the output spectra when the input signal deviates from the eigenchannel of the S-matrix, exhibiting substantial deviations from the CPA effect, which verifies the dependence on relative amplitude for complete destructive interference. The experimental data for normalized transmittance at zero detuning is illustrated in Fig. 3a, with points ABCDE corresponding to their respective spectra in Fig. 3b. The obtained results exhibit a remarkable alignment with the theoretical predictions.

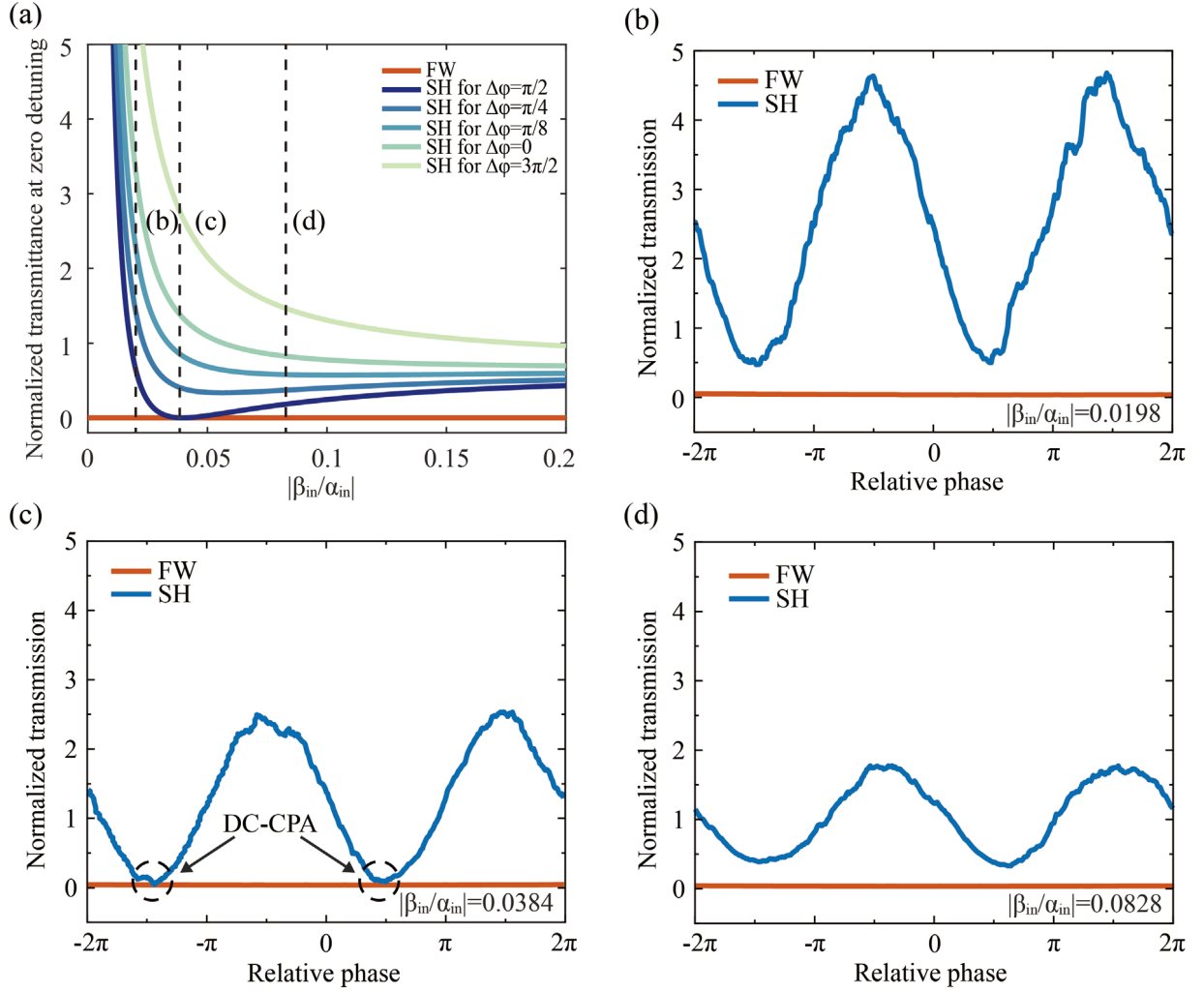

**Figure 4. Coherent control of dual-color CPA.** (a) A theoretical plot of normalized transmittance at zero detuning as a function of relative input amplitude, which is simulated under various phase differences. The absorption reaches its highest and lowest values when the relative phase is tuned to be $\pi/2$ and $3\pi/2$ respectively. (b)-(d) Experimentally obtained transmission rate of FW and SH at zero detuning as a function of relative phase, which are measured under different relative input amplitude labeled in (a). The corresponding $|\beta_{in}/\alpha_{in}|$ equals to 0.0198, 0.0384, 0.0828 respectively. DC-CPA occurs at the marked points in (c).

The absorption behavior of DC-CPA is highly sensitive to the relative phase of the inputs, as evidenced by the theoretical findings illustrated in Fig. 4a. It is apparent that variations in the phase difference result in significant shifts in the transmittance plot of SH wave versus relative amplitude. Specifically, maximal absorption occurs when the relative phase equals $\pi/2$, corresponding to total destructive interference, while minimal absorption is observed for a phase difference of $3\pi/2$, associated with constructive interference.

To investigate the behavior of phase modulation, we conducted measurements on the normalized transmittance for both the FW and SH modes around the zero-detuning point as a function of the relative phase $\Delta\varphi$. While the FW remains unaffected, the SH signal exhibits sinusoidal oscillations in phase with $\Delta\varphi$, resembling an optical phase modulator. This modulation effect is experimentally depicted in Fig. 4b-d at different relative amplitude $|\beta_{in}/\alpha_{in}| = 0.0198$, 0.0384 and 0.0828, respectively. Apparently, smaller values of $|\beta_{in}/\alpha_{in}|$ yield larger modulation depths. It is crucial to emphasize that DC-CPA only occurs at the designated point in Fig. 4c, which aligns with the pertinent eigen-channel of the S-matrix. The transmittance can be modulated over a wide range, indicating the remarkable phase sensitivity exhibited by this DC-CPA system.

In summary, we have successfully demonstrated the dual-color coherent perfect absorption in a single microcavity with second-order nonlinearity, where the FW mode and SH mode can be perfectly absorbed simultaneously. This significant advancement expands the scope of research on coherent perfect absorption into the nonlinear regime, enabling coherent control of two signals at distinct wavelengths. Given that this absorption behavior is highly sensitive to both the power and relative phase of the input waves, our system holds great promise for constructing an all-optical modulator in a cost-effective and compact form. Furthermore, apart from the sum frequency generation process depicted in our study, its reverse process known as optical parametric oscillation or spontaneous down conversion (SPDC) also exhibits significant potential for implementing a similar CPA system using the same microcavity as ours [21]. Furthermore, SPDC-associated photon-pairs generation may also be beneficial for the quantum version of CPA [31]. These findings pave the way for numerous applications including all-optical computing and quantum information processing, thereby positioning our work as a valuable contribution to these fields.

**Acknowledgment**: This work was supported by the National Science Foundation of China (Grant No. 12274295, No. 92050113); the National Key Research and Development Program (Grant No. 2023YFB3906400, No. 2023YFA1407200).